\documentclass[pra,twocolumn,preprintnumbers,amsmath,amssymb,nopacs,nofootinbib,nofootinbib,floatfix]{revtex4}

\usepackage{graphicx, bm, tikz, mathrsfs}
\usepackage[breaklinks]{hyperref}

\makeatletter
\def\graphicscale{\twocolumn@sw{0.3}{0.4}}
\def\graphicthreescale{\twocolumn@sw{0.3}{0.4}}

\begin{document}

\title{Out-of-equilibrium quantum dynamics of fermionic gases \\ in
  the presence of localized particle loss}

\author{Francesco Tarantelli}
\affiliation{Dipartimento di Fisica dell'Universit\`a di Pisa and INFN,
        Largo Pontecorvo 3, I-56127 Pisa, Italy}

\author{Ettore Vicari} 
\affiliation{Dipartimento di Fisica dell'Universit\`a di Pisa
        and INFN, Largo Pontecorvo 3, I-56127 Pisa, Italy}

\date{\today}

\begin{abstract}
  We address the effects of dissipative defects giving rise to a
  localized particle loss, in one-dimensional non-interacting lattice
  fermionic gases confined within a region of size $\ell$.  We
  consider homogeneous systems within hard walls and inhomogeneous
  systems where the particles are trapped by space-dependent external
  potentials, such as harmonic traps. We model the dissipative
  particle-decay mechanism by Lindblad master equations governing the
  time evolution of the density matrix. The resulting quantum dynamics
  is analyzed in protocols starting from the ground state of the
  Hamiltonian for $N_0$ particles, then evolving under the effect of
  one dissipative particle-loss defect, for example at the center of
  the system.  We study the interplay between time, size $\ell$ and
  the number $N_0$ of initial particles, considering two different
  situations: (i) fixed number $N_0$ of initial particles; (ii) fixed
  ratio $N_0/\ell$, corresponding to the {\em thermodynamic} limit of
  the initial equilibrium state. We show that the quantum evolutions
  of the particle number and density develop various intermediate and
  asymptotic dynamic regimes, and nontrivial large-time states when
  the dissipative mechanism acts at the center of the system.
\end{abstract}

\maketitle

% ========================= BODY =========================

\section{Introduction}
\label{intro}

The progress in atomic physics and quantum technologies allows us to
deepen our comprehension of the coherent quantum dynamics of the
many-body systems~\cite{Bloch-08,GAN-14}, and its interplay with
dissipative mechanisms arising from the interaction with the
environment~\cite{HTK-12,MDPZ-12,RDBT-13,CC-13,AKM-14,Daley-14,SBD-16},
either due to unavoidable incoherent mechanisms, or suitably
engineered system-bath couplings.  This opens the road to the
investigation of dissipation-driven phenomena, which may show the
emergence of new collective behaviors, such as novel quantum phases
and phase transitions driven by
dissipation~\cite{Hartmann-16,NA-17,MBBC-18,LCF-19,SRN-19}, peculiar
behaviors in the low-dissipative
regime~\cite{YMZ-14,YLC-16,NRV-19,RV-19,RV-20,DRV-20,RRBM-21,AC-21-2},
for example dynamic scaling at quantum transitions~\cite{RV-21-rev}.
Several studies have also addressed localized dissipative mechanisms
in many-body systems, see
e.g. Refs.~\cite{LFO-06,PP-08,Prosen-08,WTW-08,WTHKGW-11,
  KH-12,KWW-13,BGZ-14,KWBVBKW-15,Znidaric-15,VCG-18,FCKD-19,
  LHFHCE-19,TFDM-19,BMA-19,KMS-19,SK-19,SK-20,WSDK-20,FMKCD-20,
  MS-20,DMSD-20,ABWRB-20,Rossini-etal-21,TV-21,AC-21,NRG-21,
  WLGKO-09,LSHWO-15,LSHO-16}.

In this paper we address the effects of localized particle-loss
defects in lattice fermionic gases confined in a space region of size
$\ell$. We consider quantum systems where particles are
  constrained within hard walls, and systems where they are confined
  by a space-dependent potential, such as an effective harmonic
  potential. These particle systems are relevant for experimental
  setups, for example in cold-atom experiments~\cite{BDZ-08}, where a
  finite number of atoms is confined within a limited spatial region,
  typically by effectively harmonic potentials, or by effective hard
  walls, as experimentally realized in Ref.~\cite{GSGSH-13}. Within
  this class of confined systems, we investigate the dynamic features
  arising from localized particle-loss dissipative mechanisms, which
  may be controllable or inevitably present in the experiments.

  For this purpose, we consider dynamic finite-size scaling (FSS)
  frameworks, which have been particular effective to the
  investigation of thermal and quantum phase transitions, see
  e.g. Refs.~\cite{Privman-book,Cardy-book,PV-02,RV-21-rev}.  This
  approach allows us to characterize the dynamics in the presence of
  localized particle-loss dissipation, identifying different dynamic
  regimes related to different features of the dissipative system,
  whose time scales are associated with powers of the size $\ell$.  To
  achieve quantitative information on the behavior of particle systems
  of finite (relatively large) size, we consider two different
  large-size FSS limits: (i) the dynamic FSS limit keeping the initial
  number $N_0$ of particles fixed (before activating the particle-loss
  mechanism); (ii) the case in which the ratio $N_0/\ell^d$ (for $d$
  spatial dimensions) is kept fixed, corresponding to a nonzero
  chemical potential. The corresponding scaling laws provide
  complementary information on the actual dynamics of finite-size
  systems with a relatively large size.

\begin{figure}[!b]
  \includegraphics[width=1.0\columnwidth]{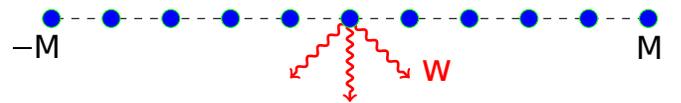}
  \caption{Sketch of a fermionic chain of size $L=2M+1$ subject to a
    localized particle loss at the central site, with strength
    controlled by the dissipation parameter $w$.  }
  \label{fig:sketch}
\end{figure}

As paradigmatic models we consider one-dimensional lattice models of
non-interacting spinless fermionic gases, within hard-wall and
harmonic traps, subject to dissipative perturbations that give rise to
a particle loss localized at one of the sites of the lattice, such as
the set up sketched in Fig.~\ref{fig:sketch}.  We model the
dissipative particle-decay mechanism by Lindblad master equations
governing the time evolution of the density
matrix~\cite{Lindblad-76,GKS-76,BP-book,RH-book,DR-21}.  To
investigate the effects of the localized particle-loss dissipation, we
study the quantum dynamics arising from protocols starting from the
ground state of the fermionic gas, then evolving under the effect of
the particle-loss dissipation, for example localized at the center of
the system.  This is analyzed in the large-$\ell$ limit for two
different initial conditions: fixed number $N_0$ of initial particles
and fixed ratio $N_0/\ell$, which corresponds to the {\em
  thermodynamic} limit of the initial fermionic gases at equilibrium,
in both hard walls and harmonic traps.  The quantum evolution of the
particle number and space-dependent density turns out to develop
various dynamic scaling regimes, and nontrivial large-time behaviors
when the dissipative mechanism acts at the center of the system.

Some issues concerning the behavior of fermionic gases in the presence
of localized dissipative interactions have been already discussed in
Refs.~\cite{FCKD-19,LHFHCE-19,KMS-19,WSDK-20,FMKCD-20,AC-21}, mainly
for homogeneous systems neglecting boundary effects. Here we extend
these studies by analyzing the interplay between time, size of the
system, and number $N_0$ of initial particles. We analyze the various
large-time and intermediate dynamic regimes, in both homogeneous
particle systems within hard walls and inhomogeneous particle systems
in harmonic traps. Substantially different and peculiar behaviors are
observed in fermionic gases confined by hard walls and harmonic traps.

The understanding of the interplay between the time dependence and the
finite size of the system is essential to interpret results in various
experimental contexts. For example this issue is fundamental for
small-size quantum simulators operating on a limited amount of quantum
objects, in the presence of controlled dissipation. We also mention
experiments with cold atoms within a trap of finite size, when the
many-body correlations become eventually sensitive to the trapping
potential (local density approximations generally fail to describe
quantum correlations, in particular when they are
critical~\cite{CTV-13} and/or out-of-equilibrium~\cite{RV-21-rev}).

The paper is organized as follows. In Sec.~\ref{modeldiss} we present
the models of spinless fermionic gases confined within hard walls or
traps arising from external power-law potentials, the description of
the localized particle-loss dissipation, the dynamic protocols that we
consider to study the dissipative effects, and the main equations that
allow us to study the evolution of the particle number and density.
In Sec.~\ref{hwbc} we present results for fermionic gases confined
within hard walls, discussing the main features of the asymptotic
stationary states, and the emerging dynamic scaling behaviors
characterizing the quantum evolution arising from the protocols. In
Sec.~\ref{hartra} we focus on fermionic gases within harmonic traps,
showing again the emergence of various dynamic regimes. Finally, in
Sec.~\ref{conclu} we summarize and draw our conclusions.

\section{Free lattice fermions  with localized dissipative defects}
\label{modeldiss}

\subsection{The Hamiltonian}
\label{model}

We consider one-dimensional $N$-particle Fermi gases defined on a
chain with $L=2M+1$ sites, by the Hamiltonian
\begin{equation}
  \hat H =
  - \kappa \sum _{x=-M}^{M-1} 
  (\hat c_{x}^\dagger \hat c_{x+1} +  \hat c_{x+1}^\dagger \hat c_{x})\,,
  \label{Hfree}
\end{equation}
where $\hat c_x$ is a fermion one-particle operator, and $\hat n_x =
\hat c_x^\dagger c_x$ is the particle density operator.  We consider
hard-wall (open) boundary conditions.  The site $x=0$ is the central
site of the chain.  In the following we set $\hslash=1$, $\kappa=1$,
and the lattice spacing $a=1$, from which one can easily derive the
units of all other dimensionful quantities considered in the rest of
the paper.

We also consider fermionic systems where the particles are trapped by
an external potential, which can be taken into account by adding a
corresponding term to the Hamiltonian (\ref{Hfree}), such as
\begin{eqnarray}
  \hat H_t =
  - \sum _x 
  (\hat c_{x}^\dagger \hat c_{x+1} +  \hat c_{x+1}^\dagger \hat c_{x})
+  \sum_x V(r) \, \hat n_x\,,\label{htrap}
  \end{eqnarray}
where 
\begin{eqnarray}
\hat n_x =
\hat c_x^\dagger c_x\,,\quad  V(r)= (r/L_t)^p\,, \quad r\equiv |x|\,,
   \label{potential}
\end{eqnarray}
$p$ is a positive number, $r$ is the distance from the center $x=0$ of
the trap, and $L_t$ plays the role of trap size.~\footnote{$L_t$ plays
the role of trap size~\cite{BDZ-08,RM-04,CV-10-2}, so that the {\em
  thermodynamic} limit is obtained in the large trap-size limit,
$L_t\to \infty$, keeping the ratio between the particle number $N$ and
the trap size $L_t$ constant, which can be equivalently obtained by
adding a chemical-potential term in the Hamiltonian, such as the one
reported in Eq.~(\ref{chemicalpot}).}  The trapping potential is
effectively harmonic in most cold-atom experiments~\cite{BDZ-08},
i.e., $p=2$.  In the limit $p\to\infty$ we recover the model
(\ref{Hfree}) with hard-wall boundary conditions and $M=\lfloor L_t
\rfloor$.  The size of systems described by the Hamiltonian $\hat H_t$
with finite $p$ is supposed to be infinite. However for practical
purposes it is sufficient to consider models within hard walls with $L
\gg L_t$. Indeed the large-size convergence is generally fast for
sufficiently large values of $p$, including $p=2$, due to the fact
that the average particle density $\langle \hat{n}_x \rangle$ vanishes
rapidly for $|x|\gg L_t$. The main features of the behavior of
fermionic gases trapped by a inhomogeneous external power-law
potentials have been much investigated, see e.g.
Refs.~\cite{ACV-14,Nigro-17,CV-10,CV-10-2,CV-10-3,CV-10-4,Pollet-12,V-12,CLM-15}.

In systems within both hard-wall and inhomogeneous traps, the particle
number operator
\begin{equation}
  \hat{N} = \sum_x \hat n_{x}\,
\label{partnum}
  \end{equation}
commutes with both Hamiltonians (\ref{Hfree}) and
(\ref{htrap}). Therefore the particle number is conserved in both
cases.  In the following we consider ground states for a number $N_0$
of particles as starting point of dynamic protocols involving
dissipative mechanisms.

\subsection{Localized particle-decay dissipation}
\label{locdiss}

We model the dissipative mechanisms within the Lindblad
framework~\cite{Lindblad-76,GKS-76}, where the evolution of the matrix
density $\rho(t)$ of the system is described by he
equation~\cite{BP-book,RH-book}
\begin{eqnarray}
&&  {\partial\rho\over \partial t} = {\cal L}[\rho]=
  - i \, [ \hat H,\rho]
  + {\mathbb D}[\rho]\,.
  \label{EQLindblad}
\end{eqnarray}
We recall that the conditions leading to the Lindblad framework are
typically satisfied in quantum optical
implementations~\cite{SBD-16,DR-21}.  The form of the operator
${\mathbb D}[\rho]$ depends on the nature of the dissipation arising
from the interaction with the bath.  We consider a localized
particle-decay dissipation acting at the site $z$, modeled by the
Lindblad operator~\cite{HC-13, KMSFR-17, Nigro-19, NRV-19, WSDK-20,
  FMKCD-20, DR-21,RV-21-rev}
\begin{eqnarray}
\mathbb{D}[\rho] = w\,\biggr[
    \hat c_{z}\,\rho\,\hat c_{z}^\dagger - {1\over 2}\left( \rho\,
    \hat c_z^\dagger \hat c_{z} + \hat c_z^\dagger \hat c_{z} \rho \right)
    \biggr] \,, 
\label{Lindop}
\end{eqnarray}
where $w$ is a parameter controlling the strength of the particle-loss
dissipation.

The reflection symmetry with respect to the center of the confined
particle system is only preserved when the particle loss is localized
at the center. As we shall see, this will lead to peculiar behaviors
with respect to the case of particle-loss dissipation localized at
generic sites.

The approach to the asymptotic stationary states are generally
controlled by the Liouvillian gap $\Delta_{\cal L}$ associated with
the generator ${\cal L}$ entering the Lindblad equation
(\ref{EQLindblad})~\cite{BP-book,RH-book,Znidaric-15,MBBC-18,SK-20}.
The asymptotic stationary state is provided by the eigenstate of
${\cal L}$ with vanishing eigenvalue, $\Lambda_0=0$, while all other
eigenstates have eigenvalues $\Lambda_i$ with negative real part,
i.e. ${\rm Re}\,\Lambda_i<0$ for any $i>0$.

\subsection{Dynamic protocol}
\label{dynprot}

To study fermionic gases under the effects of a localized
particle-loss mechanism, we consider the following dynamic protocol,
for systems within both hard-wall and harmonic traps, respectively of
size $L=2M+1$ and $L_t$.

\begin{itemize}

\item[$\bullet$] The protocol starts at time $t=0$ from the ground
  state of the Hamiltonian (\ref{Hfree}) or (\ref{htrap}) with a
  number $N_0$ of particles. We recall that the ground state of $N_0$
  noninteracting fermionic particles is obtained by filling the lowest
  $N_0$ one-particle energy levels.

\item[$\bullet$] The time evolution for $t>0$ is driven by the 
  Lindblad equation (\ref{EQLindblad}) for the density matrix
  $\rho(t)$, with particle-decay dissipation localized at a site $z$
  and controlled by the parameter $w$.

\item[$\bullet$]
The particle density and total particle number,  
\begin{eqnarray}
n_x(t) = {\rm Tr}[ \rho(t) \hat n_x ] \,,\qquad
N(t) = {\rm Tr}\Bigr[ \rho(t) \hat N \Bigr] \,,
\label{nxntdef}
\end{eqnarray}
are monitored during the out-of-equilibrium evolution for $t>0$, up to
their large-time behaviors.

\end{itemize}

To compute the particle density $n_x(t)$ and particle number $N(t)$,
we proceed as follows.  We introduce the correlation functions
\begin{eqnarray}
  {\mathscr C}_{x,y}(t)= {\rm Tr}\Bigr[ \rho(t) \, \hat c_x^\dagger
    \hat c_y \Bigr] \,.
\label{defct}
\end{eqnarray}
For homogeneous systems described by the Hamiltonian (\ref{Hfree}),
the Lindblad equation (\ref{EQLindblad}) 
implies
\begin{eqnarray}
  {d\mathscr{C}_{x,y}\over dt} &=& i\,( \mathscr{C}_{x,y+1} -
   \mathscr{C}_{x-1,y} + \mathscr{C}_{x,y-1} - \mathscr{C}_{x+1,y} )
      \nonumber\\
    &&- \frac{w}{2} \ ( \delta _{z,y} + 
   \delta _{x, z} ) \, \mathscr{C}_{x,y} \,,
   \label{eqscxy}
\end{eqnarray}
where $\delta_{x,x}=1$ and $\delta_{x,y}=0$ for $x\neq y$.  Since we
consider open (hard-wall) boundary conditions, ${\mathscr
  C}_{xy}(t)=0$ when the coordinates $x$ or $y$ refer to sites outside
the space interval $[-M,M]$.  An analogous equation can be derived in
the presence of an inhomogeneous external trapping potential,
cf. Eq.~(\ref{htrap}).  We obtain
\begin{eqnarray}
  &&  
  {d\mathscr{C}_{x,y}\over dt} = i\,( \mathscr{C}_{x,y+1} -
   \mathscr{C}_{x-1,y} + \mathscr{C}_{x,y-1} - \mathscr{C}_{x+1,y} )
   \nonumber\\
&&\;\; + i {|x|^p - |y|^p\over L_t^p} \mathscr{C}_{x,y} 
- \frac{w}{2} \ ( \delta _{z,y} + 
   \delta _{x, z} ) \, \mathscr{C}_{x,y} \,.
   \label{eqscxytrap}
\end{eqnarray}
Then, after numerically solving the above equations, we use the
relations
\begin{equation}
n_x(t)=\mathscr{C}_{x,x}(t) \,,\qquad   N(t) = \sum _x n_x\,.
\label{ntcxy}
\end{equation}

One can easily check that for both hard-wall and harmonic traps the
derivative of the particle number is proportional to the average
particle density $n_z$ at the site $z$ where the particle-decay
dissipation is localized, i.e.
\begin{equation}
  {d N(t)\over dt} = - w \, n_z(t) < 0\,.
\label{ntnz}
\end{equation}
Therefore the particle number decays monotonically, since $n_z(t)\ge
0$, and the particle loss stops if $n_z(t)= 0$ asymptotically.

One may also consider the energy of the system, defined as
\begin{eqnarray}
E(t) = {\rm Tr}[\rho(t) \hat H] \,,
\label{enedef}
\end{eqnarray}
for which the Lindblad equation implies
\begin{eqnarray}
  {d E(t)\over dt} = {\rm Tr}\left[{d\rho(t)\over dt} \hat H\right]= w
  {\rm Tr}[{\mathbb D}[\rho] \,\hat H]\,.
  \label{detg}
\end{eqnarray}
For systems with particle-loss dissipation localized at the central
site $x=0$, we obtain
\begin{eqnarray}
  {d E(t)\over dt} =
  w \, {\rm Re}\,({\mathscr C}_{0,1}+{\mathscr C}_{-1,0}) =
  2 w \, {\rm Re}\,{\mathscr C}_{0,1}\,,
  \label{derene}
\end{eqnarray}
which holds for systems within hard walls and also inhomogeneous
traps.

\section{Fermi gases within hard walls}
\label{hwbc}

In this section we consider homogeneous Fermi chains, cf.
Eq.~(\ref{Hfree}), and discuss the dynamic evolution under the
particle-loss dissipation described by the Lindblad equation
(\ref{EQLindblad}), in particular Eq.~(\ref{eqscxy}). For this
purpose, we numerically solve the differential equation (\ref{eqscxy})
using the fourth-order Runge-Kutta method (with an accuracy of
approximately $10^{-8}$ on the evolution of the particle number).

We study the interplay between the time dependence, the number $N_0$
of initial particles and the size $L=2M+1$ of the lattice.  For this
purpose we consider two different situations: (i) the number $N_0$ of
particles is kept fixed while increasing $M$; (ii) the number of
particles is increases as $N_0\sim M$, so that the ratio $N_0/M$
fixed, while increasing $M$. Note that the latter condition can be
equivalently realized in the large-size limit by adding a chemical
potential to the Hamiltonian (\ref{Hfree}), i.e.
\begin{equation}
\hat H_\mu = -(\mu+2) \sum _x \hat n_x \,.
\label{chemicalpot}
\end{equation}
The value $\mu=\mu_{\rm vs}=-2$ corresponds to the vacuum-superfluid
transition point~\cite{Sachdev-book,ACV-14}, separating the phase
where the lowest Hamiltonian eigenstate has $N=0$ particles from the
one for $\mu>-2$ where the ground-state has $N\sim L$ fermions.

In the following we first analyze the asymptotic large-time regime.
Then we show that the time dependence of the particle number develops
various asymptotic and intermediate regimes, which may differ in the
cases we keep $N_0$ or $N_0/M$ fixed.

\subsection{Asymptotic stationary states}
\label{asysta}

For generic locations of the particle-loss defects, the asymptotic
stationary state turns out to be trivial, i.e. an empty state without
particles. However in some cases, in particular when the defect is
localized at the center of the chain, the quantum evolution of the
system keeps a residual number of particles even in the large-time
limit.

This can be shown analytically, straightforwardly extending the
analysis for non-interacting bosons reported in Ref.~\cite{KH-12}, to
free fermions.  Since we are considering systems of size $L=2M+1$ with
hard-wall boundary conditions, we introduce the fermionic operators
\begin{eqnarray}
  \hat \eta_k = \sqrt{2\over L+1}
  \sum_{y=1}^{L} \sin\left({\pi k y \over L+1}\right)
  \hat c_y\,, \label{trafou}
%  \hat c_y = \sqrt{2\over L+1} \sum_{k=1}^{L} \sin\left({\pi k y
%  \over L+1}\right) \hat \eta_k \,,
\end{eqnarray}
where, to simplify the formulas, we have shifted the coordinates so
that $y = x + M+1$ (therefore the site coordinates are $y=1,...,L$,
and the center is located at $y=M+1$).  This allows us to write the
Hamiltonian (\ref{Hfree}) as
\begin{equation}
  \hat H = - 2 \sum_{k=1}^L \cos\left({\pi k\over L+1}\right) \hat n_k\,,
  \qquad   \hat n_k = \hat \eta_k^\dagger \hat \eta_k\,.
  \label{Hfreek}
  \end{equation}
%Using the relations (\ref{trafou}), the Lindblad
%operator (\ref{Lindop}) can be written as
%\begin{eqnarray}
%  \mathbb{D}[\rho] &=& {w\over L+1} \sum_{q,q'=1}^L \sin\left({\pi q z
%    \over L+1}\right) \sin\left({\pi q' z \over L+1}\right)
%  \nonumber\\ && \times \left( 2 \hat \eta_{q'} \rho \hat \eta_q^\dagger
%  - \hat \eta_q^\dagger \hat \eta_{q'} \rho - \rho \hat \eta_q^\dagger \hat
%  \eta_{q'} \right)\,.
%    \label{drhokappa}
%\end{eqnarray}
The operator $\hat n_k$ commutes with the Hamiltonian, i.e. $[\hat
  H,\hat n_k] = 0$, and satisfies $\sum_k \hat n_k = \hat N = \sum_x
\hat n_x$.  Its expectation value 
\begin{equation}
n_k(t) = {\rm Tr}[\rho(t)\, \hat n_k]\,
\label{nkope}
\end{equation}
counts the number of particles associated with the mode $k$.  The
initial equilibrium ground state with $N_0$ fermionic particles is
constructed by filling the first $N_0$ one-particle energy levels,
thus at $t=0$ we have $n_k=1$ for $k\le N_0\le L$, and zero
otherwise. The modes with odd (even) $k$ are even (odd) under
inversion with respect to the center $y=M+1$ of the chain.  The time
evolution of $n_k$ is determined by the Lindblad equation for the
density matrix.  Considering a particle-decay dissipation located at a
generic site $z$, straightforward calculations lead to the equation
\begin{eqnarray}
  {d n_k\over dt} &=&
%  - {w\over M+1} \sum_{q,q'=1}^M \sin\left({\pi q z
%  \over M+1}\right) \sin\left({\pi q' z \over M+1}\right)
%\nonumber\\ &&\times (\delta_{kq} + \delta_{kq'}) {\rm Tr}[\rho(t) \,
%  \hat \eta_q^\dagger \hat \eta_{q'}]\,
%\nonbumber \\
- {w\over L+1} \sin\left({\pi k z \over L+1}\right)
\label{ddtnk} \\
&\times& \sum_{q=1}^L \sin\left({\pi q z\over L+1}\right)
{\rm Tr}[\rho(t) \,
(\hat \eta_q^\dagger \hat \eta_{k} + \eta_k^\dagger \hat \eta_{q})]\,,
\nonumber
\end{eqnarray}
where, due to the fact that $[\hat H,\hat n_k]=0$, the only
contribution to the time derivative of $n_k$ comes from the
dissipative term.  We note that the r.h.s. of Eq.~(\ref{ddtnk})
vanishes when
\begin{equation}
k z = j (L+1)\,, \quad j=1,2,...,
\label{kz}
\end{equation}
thus implying the conservation of the corresponding particle number
$n_k$ even in the presence of localized particle-decay dissipation.

\begin{figure}[!t]
  \includegraphics[width=0.95\columnwidth]{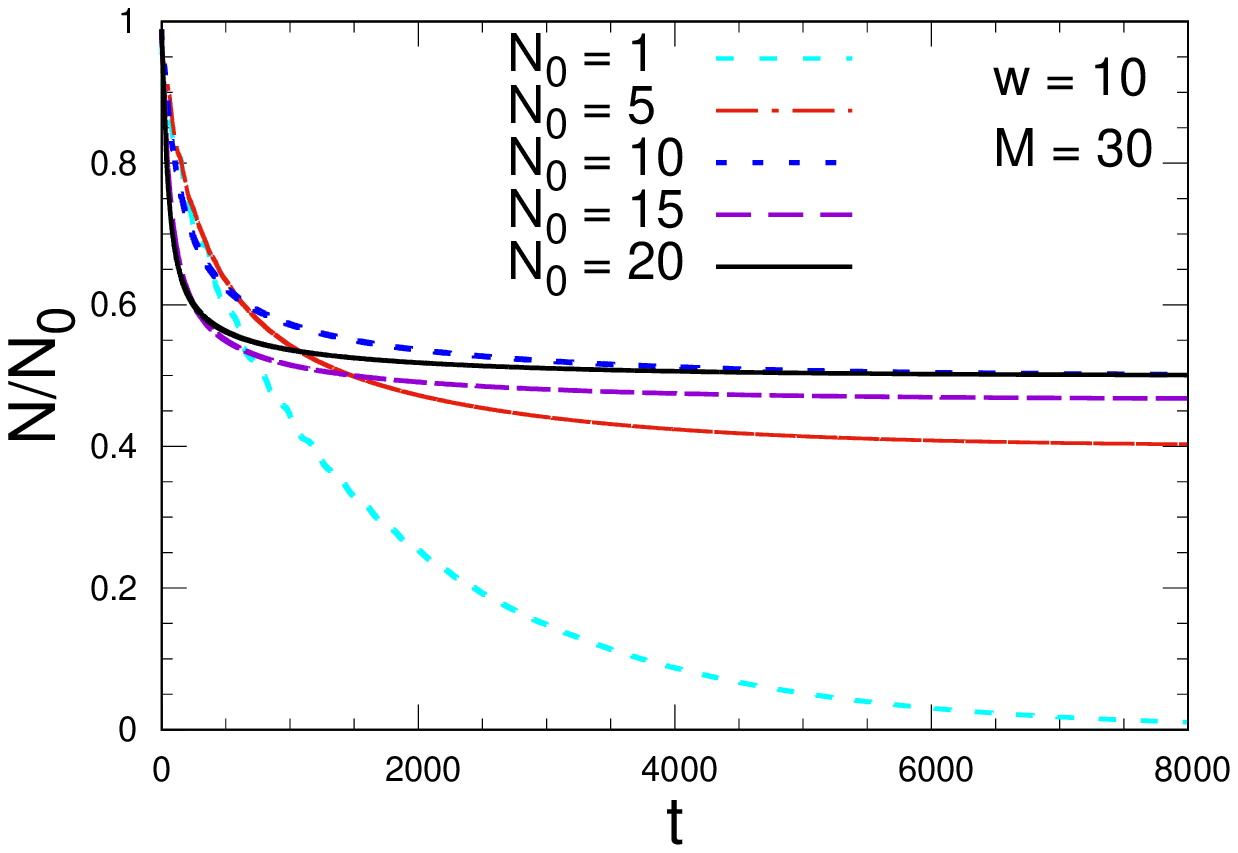}
  \includegraphics[width=0.95\columnwidth]{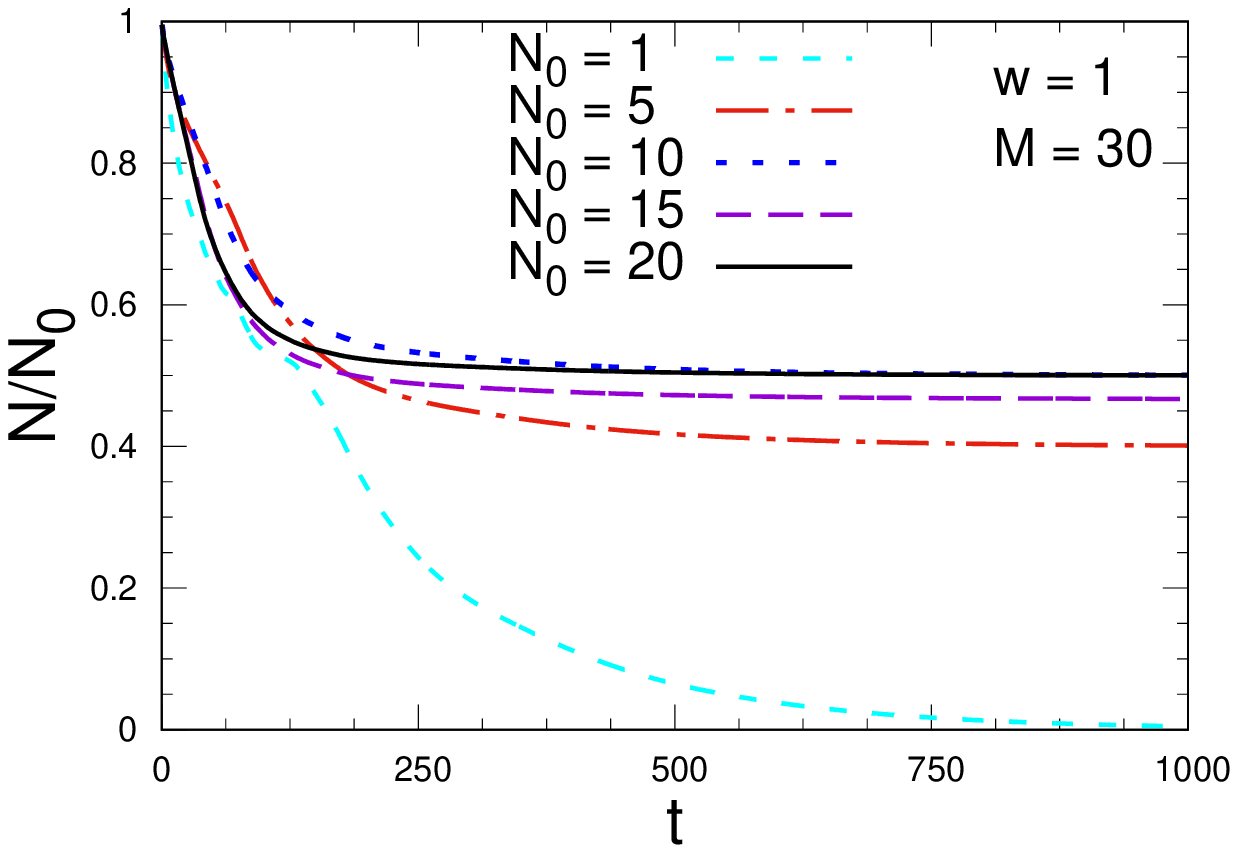}
  \caption{Behavior of the ratio $N(t)/N_0$ for the central-site
    particle-loss dissipation in systems of size $L=61$ ($M=30$)
    within hard walls, various initial particle number $N_0$, and
    dissipation localized at the center of the system, with $w=1$
    (bottom) and $w=10$ (top).  In both cases asymptotic stationary
    limit turns out to converge to $N/N_0=1/2$ for even $N_0$, and
    $N/N_0=(N_0-1)/(2N_0)$ for odd $N_0$.  Note that approach to the
    asymptotic value is slower for $w=10$ than $w=1$.}
  \label{ndiffn0}
\end{figure}

If we consider a central-site dissipation, thus $z=M+1=(L+1)/2$ [we
  recall that we are using shifted coordinates with respect to
  Eq.~(\ref{Hfree})], then the condition (\ref{kz}) reduces to $k=2j$,
thus implying that $n_k$ remains unchanged for all even $k$ (whose
odd-parity modes vanish at the central dissipative site), while it
gets suppressed for odd $k$. Therefore, Eq.~(\ref{ddtnk}) implies that
half of the fermions survives centrally localized decay
dissipation. More precisely the stationary states are characterized by
a residual particle number $N_{\rm asy} = N_0/2$ for even $N_0$, and
$N_{\rm asy} = (N_0-1)/2$ for odd $N_0$.

Note that the particle loss localized at the center, preserving the
parity symmetry with respect to the center of the chain, is the
optimal one to keep a fraction of fermionic particles at large time.
For example, in the case of a dissipation at the boundaries, i.e.,
when $z=1$ or $z=M$ in Eq.~(\ref{ddtnk}), no particles survive because
all $k$-modes are involved by the Lindblad operator, leading to the
complete suppression of the particles filling the initial ground
state.

\begin{figure}[!t]
  \includegraphics[width=0.95\columnwidth]{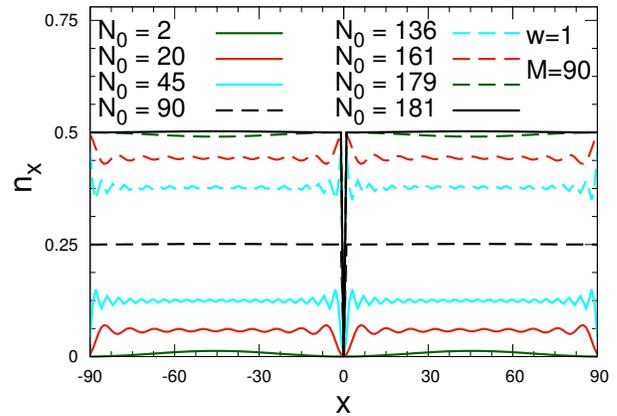}
  \caption{Data for the quantum evolution of the fermionic gas within
    hard walls, size $L=2M+1$ with $M=90$, initial particle number
    $N_0=10$, dissipation localized at the center of the chain with
    $w=1$. We show the particle density $n_x(t)$ at the sites $x=0$
    and $x=10$ (top), and the ratio $N(t)/N_0$ (bottom).  They
    approach asymptotic stationary limits (at least within the
    numerical precision, which is very accurate).}
  \label{nxdiffn0time}
\end{figure}

\begin{figure}[!t]
  \includegraphics[width=0.95\columnwidth]{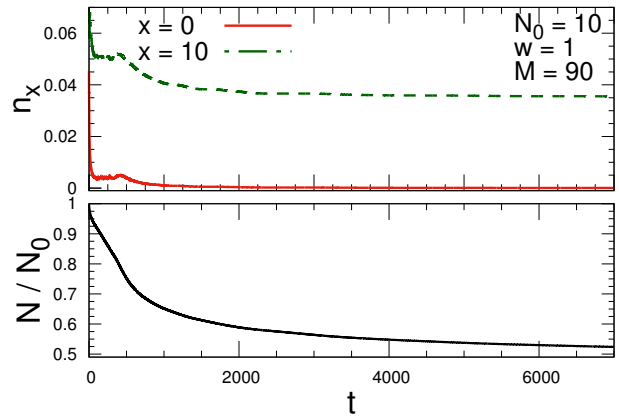}
  \caption{Asymptotic stationary limit of the particle density $n_x$
    for systems of size $L=181$ ($M=90$) within hard-wall boundary
    conditions, in the presence of a central-site dissipation with
    $w=1$, and for various $N_0$. In all cases the particle density
    vanishes for $x=0$, and it is almost flat elsewhere, apart from
    small spatial oscillations, which appear similar to the
      Friedel oscillations characterizing the behavior of closed
      particle systems.  }
  \label{nxdiffn0}
\end{figure}

The above analytical results are confirmed by the numerical results
for the central particle-loss dissipation, see for example
Fig.~\ref{ndiffn0} where we show the time dependence of the ratio
$N(t)/N_0$ for various values of $N_0$ and $w$, and in particular the
approach to its nonzero asymptotic limit. Also the space dependence of
the particle density $n_x$ turns out to become stable asymptotically,
approaching a stationary configuration, as shown in
Fig.~\ref{nxdiffn0time}.  In Fig.~\ref{nxdiffn0} we show some results
for the spatial dependence of the average particle density $n_x$ of
the asymptotic stationary states. We note that $n_x$ is quite flat
except at $x=0$ where the dissipative mechanism acts, and at the
boundaries of the chain (essentially due to the hard-wall boundary
conditions). The almost flat region shows some spatial oscillations,
which appear suppressed when $N_0\approx M$ and $N_0\approx 2M$.

\subsection{Approach to the asymptotic states}
\label{asyappro}

\subsubsection{Large-size behavior of the Liouvillian gap}
\label{liogap}

The approach to the stationary state is controlled by the Liouvillian
gap $\Delta_{\cal L}$ of the generator ${\cal L}$ of the Lindblad
equation~\cite{BP-book,RH-book,Znidaric-15,MBBC-18,SK-20},
\begin{equation}
\Delta_{\cal L} = - {\rm Max}_{i>0} \, {\rm Re}\,(\Lambda_i)\,,
\label{deltadeb}
\end{equation}
where $\Lambda_i$ are the eigenvalues of ${\cal L}$ (we recall that
the largest eigenvalue is $\Lambda_0=0$ and ${\rm Re} \,\Lambda_i<0$
for any $i>0$).  The Liouvillian gap for homogeneous spin chains and
fermionic wires with localized dissipative mechanisms, such as that
described by Eq.~(\ref{Lindop}), shows generally the asymptotic
finite-size
behavior~\cite{PP-08,Prosen-08,Znidaric-15,SK-20,TV-21}
\begin{equation}
\Delta_{\cal L}(w,L)\approx D_{\cal L}(w) \,L^{-3}\,.  
\label{deltaL}
\end{equation}
We expect that this asymptotic large-$L$ behavior holds independently
of the location of the particle-decay dissipation, and it does not
depend on the initial conditions, thus on $N_0$.  The scaling equation
(\ref{deltaL}) implies that the approach to the asymptotic behavior
becomes slower and slower with increasing the size $L$ of the lattice
at fixed $w$.

\begin{figure}[!t]
  \includegraphics[width=0.95\columnwidth]{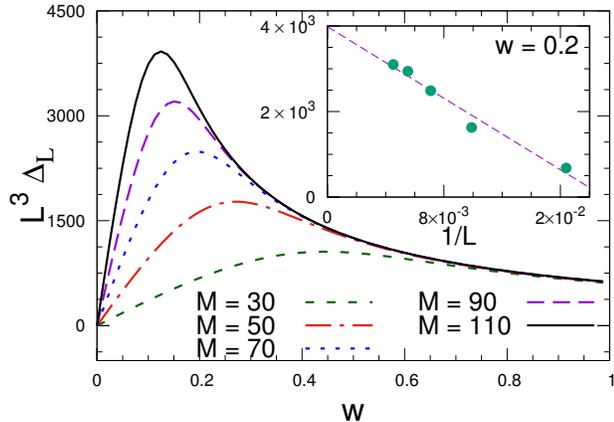}
  \caption{The Liouvillian gap $\Delta_{\cal L}$ for particle-decay
    dissipation localized at the center of the chain, for various
    system size $L=2M+1$. The curves appear to converge with
      increasing $M$; this is clearly shown at least for $w\gtrsim
      0.2$, as also shown by the large-$L$ convergence at a fixed
      value $w=0.2$ [suggesting that the corrections to the asymptotic
        scaling behavior (\ref{deltaL}) are approximately
        $O(L^{-1})$].  We conjecture that the convergence extends to any
      $w>0$, but it is nonuniform when decreasing $w$ toward zero, see
      text.  }
  \label{liogaps}
\end{figure}

\begin{figure}[!t]
  \includegraphics[width=0.95\columnwidth]{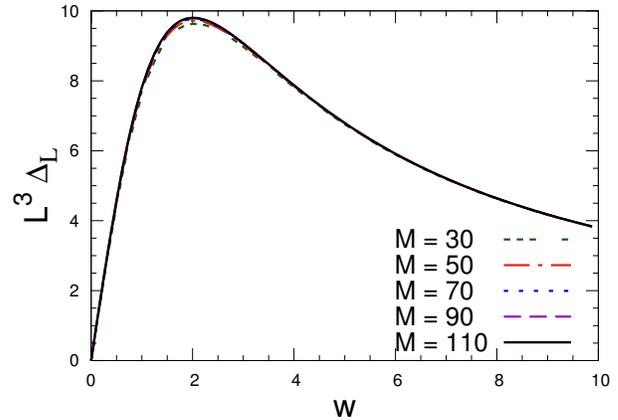}
  \caption{Scaling behavior of the Liouvillian gap $\Delta_{\cal L}$
    for particle-decay dissipation localized at one of the boundaries
    of the chain.}
      \label{liogapsb}
\end{figure}

\begin{figure}[!t]
    \includegraphics[width=0.95\columnwidth]{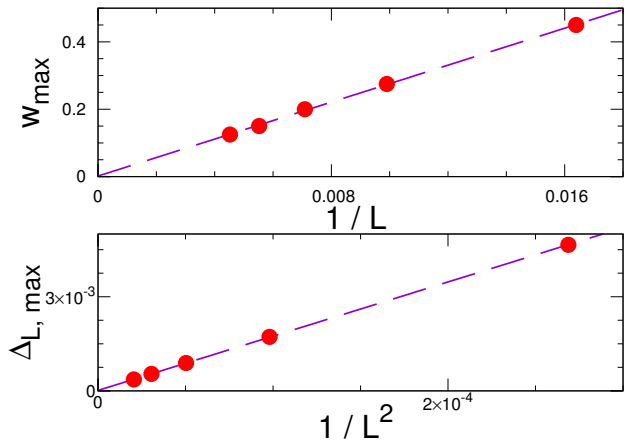}
  \caption{Some details of the behavior of the Liouvillian gap
    $\Delta_{\cal L}$ for dissipation localized at the center of the
    lattice. We show the location $w_{\rm max}$ of the maximum of the
    Liouvillian gap (top), showing that $w_{\rm max}\sim L^{-1}$, and
    value of $\Delta_{\cal L}$ at the maximum (bottom), showing that
    $\Delta_{\cal L}(w_{\rm max})\sim L^{-2}$.}
  \label{gapcenterdetails}
\end{figure}

\begin{figure}[!t]
  \includegraphics[width=0.95\columnwidth]{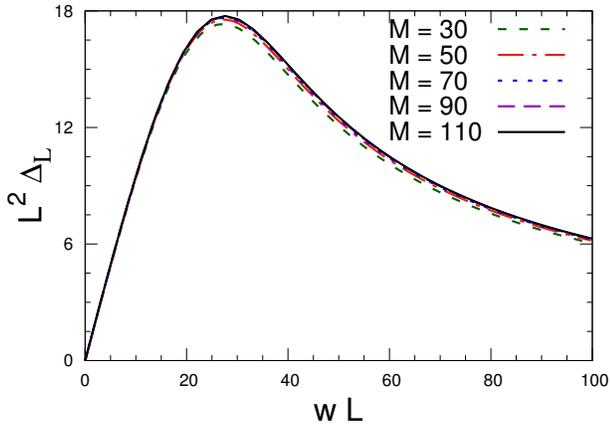}
  \caption{Plots of $L^2 \Delta_{\cal L}$ versus $wL$ for systems
    within hard walls and with central particle-loss dissipation, for
    various $L=2M+1$.  They support the scaling equation
    (\ref{deltaLcenter}).}
      \label{gapcenter2}
\end{figure}

The asymptotic behavior (\ref{deltaL}) is confirmed by numerical
analyses of the Liouvillian gap using the method outlined in
Ref.~\cite{PP-08}, see App.~\ref{appa}. Some numerical results are
shown by Figs.~\ref{liogaps} and \ref{liogapsb}, for particle-decay
dissipation localized at the center and at the boundary of the chain,
respectively.

In both cases, $\Delta_{\cal L}$ appears nonmonotonic, increasing for
small values of $w$ and decreasing for sufficiently large dissipation
strength $w$, for any $L$.  The approach to the asymptotic behavior
becomes slower and slower with increasing $w$ for large $w$. In
particular, $\Delta_{\cal L}(w,L)$ shows the large-$w$ behavior $L^3
\Delta_{\cal L}(w)\approx D_{\cal L}(w)\sim w^{-1}$. This explains the
results shown in Fig.~\ref{ndiffn0}, where the approach to the
asymptotic value of the particle number for $w=10$ turns out to be
slower than that for $w=1$.  The suppression in the limit of
  strong dissipation, for large $w$, may be interpreted as a
  quantum-Zeno-like phenomenon~\cite{MS-77,FP-08}, where a strong
  dissipation somehow slows down the dynamics, see
  e.g. Refs.~\cite{TNDTT-17,DMSD-20} for the emergence of similar
  quantum Zeno regimes.

As shown by Figs.~\ref{liogaps} and \ref{liogapsb}, the Liouvillian
gap at fixed $L$ has a maximum at an intermediate value of $w$. The
corresponding values $w_{\rm max}$ and $L^3 \Delta_{\cal L}(w_{\rm
  max})$ appears to rapidly converge to the large-$L$ for dissipations
localized at the boundaries.  The approach to the asymptotic $L^{-3}$
behavior (\ref{deltaL}) appears significantly slower in the case of
dissipation localized at the center.  The curves for different lattice
sizes appear to converge for $w\gtrsim 0.2$, see Fig.~\ref{liogaps}
and in particular its inset. We conjecture that, like the case of
dissipation at the boundaries, the convergence in the large-size limit
extends to any $w>0$, but it is nonuniform when decreasing $w$ toward
zero (i.e. the correction diverges for $w\to 0$). This is also
supported by the plots reported in Figs.~\ref{gapcenterdetails}, which
show that the location of the maximum value of the Liouvillian gap for
central dissipation moves toward $w=0$, with $w_{\rm max}\sim L^{-1}$,
and its maximum value decreases as $\Delta_{\cal L}(w_{\rm max})\sim
w_{\rm max}^2\sim L^{-2}$, instead of the general asymptotic $L^{-3}$
behavior for $w>0$ fixed. Actually, they suggest that the Liouvillian
gap for central dissipation shows also the asymptotic scaling behavior
\begin{equation}    
\Delta_{\cal L}(w,L)\approx {\cal D}_{\cal L}(wL) \,L^{-2}\,,
\label{deltaLcenter}
\end{equation}
obtained in the large-$L$ limit keeping $wL$ constant.  This scaling
behavior is demonstrated by the plot reported in
Fig.~\ref{gapcenter2}.  Therefore the function $D_{\cal L}(w)$
entering the asymptotic $L^{-3}$ behavior (\ref{deltaL}) must be
singular for $w\to 0$ in the case of central-site dissipation,
diverging as $w^{-1}$.

Note that the above considerations on the behavior of the Liouvillian
gap apply to generic values of $w$, and,  of course, they do not depend
on the initial number $N_0$ of particles.  In the following we will
mainly present results for the value $w=1$ of the dissipation
parameter, whose dynamic scenarios are shared with those arising from
generic finite values of $w$.

\subsubsection{Large-time behavior of the particle number}
\label{lasypgap}

\begin{figure}[!t]
  \includegraphics[width=0.95\columnwidth]{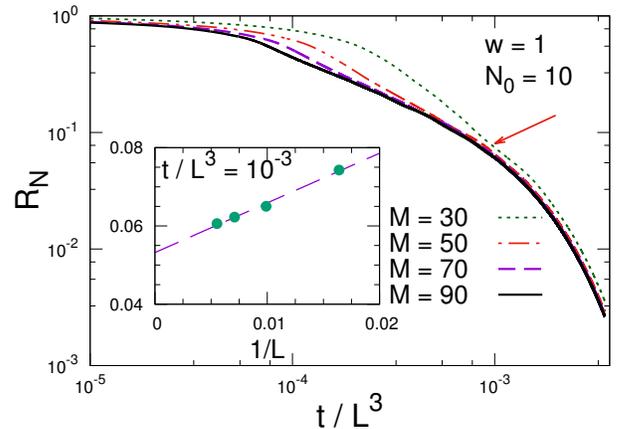}
  \caption{The time dependence of the ratio $R_N$ versus $t/L^3$
    ($L=2M+1$), for homogenous systems within hard walls with $N_0=10$
    and central particle-loss dissipation with $w=1$.  The data for
    various sizes $M$ show clearly the convergence toward a dynamic
    scaling curve, approximately as $1/L$ as shown by the inset for a
    particular value of the ratio $t/L^3$ (the one indicated by the
    arrow in the main figure).}
  \label{rntn010}
\end{figure}

On the basis of the large-size behavior of the Liouvillian gap, we
expect that the time scale $t_a$ of the approach to the stationary
state is given as
\begin{equation}
t_a \sim \Delta_{\cal L}^{-1}\sim L^3\,,
\label{timescaleasy}
\end{equation}
at fixed $w>0$. This time scale must characterize the approach to the
stationary limits of the particle number and density.  This is
confirmed by the numerical computations, see for example
Fig.~\ref{rntn010} where we report results for the ratio
\begin{equation}
  R_N(t)\equiv {N(t)-N_{\rm asy}\over N_0-N_{\rm asy}}\,,
    \label{defqu}
\end{equation}
for protocols starting from a fixed number $N_0$ of particles.  They
show the asymptotic large-$L$ scaling behavior
\begin{eqnarray}
  R_N(t,w,L) \approx A(t/L^3,w)\,,\label{scalbehasy}
\end{eqnarray}
which also implies
\begin{eqnarray}
{d R_N(t,w,L)\over dt} &\approx& L^{-3} B(t/L^3,w)\,.  \nonumber
\end{eqnarray}
As usual within FSS frameworks, the above asymptotic FSS
  behaviors are expected to hold in the large-$L$ limit keeping the
  ratio $t/L^3$ and $w$ fixed.

\begin{figure}[!t]
  \includegraphics[width=0.95\columnwidth]{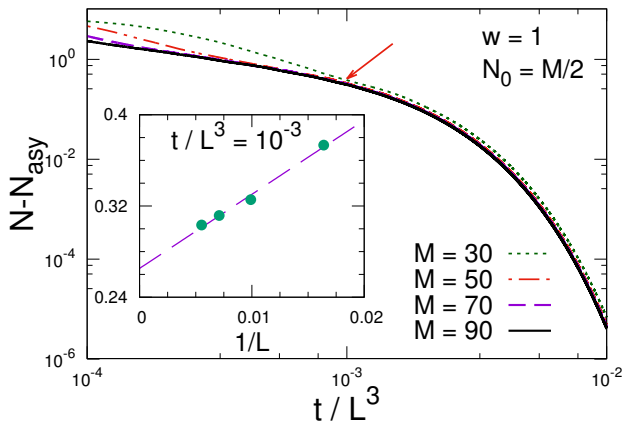}
  \caption{The difference $N(t) - N_{\rm asy}$ vs $t/L^3$, for systems
    within hard walls with $N_0/M=1/2$, and central dissipation with
    $w=1$.  Again we observe the asymptotic convergence toward a
    dynamic scaling curve, as shown by the inset for a particular
    value of $t/L^3$ (indicated by the arrow in the main figure).}
  \label{dntln0lasy}
\end{figure}

Analogous results are obtained for the case we start from a fixed
ratio $N_0/M$ (we recall that $L=2M+1$), as shown in
Fig.~\ref{dntln0lasy} for $N_0/M=1/2$, where we plot $N(t)-N_{\rm
  asy}$ versus $t/L^3$ and the curves appear to collapse in the
large-$M$ limit. Note that when $N_0/M$ is kept fixed, the quantity
$R_N$ defined in Eq.~(\ref{defqu}) is not appropriate, because it is
always suppressed in the large-time limit, due to the denominator that
behaves as $N_0-N_{\rm asy}\sim L$.

As we will show below, this is not the end of the story, indeed
further peculiar intermediate scaling behaviors emerge, differing
between the cases $N_0$ and $N_0/M$ fixed.

\subsection{Intermediate scaling behaviors keeping $N_0$ fixed}
\label{N0fixed}

\begin{figure}[!t]
  \includegraphics[width=0.95\columnwidth]{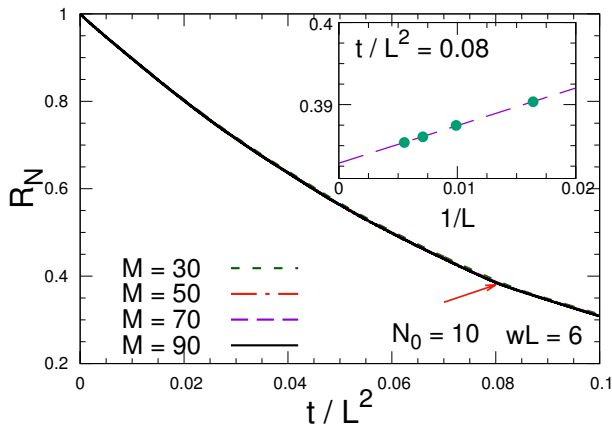}
  \caption{The ratio $R_N$ versus $t/L^2$ for systems within hard
    walls with $N_0=10$, central dissipation with $wL=6$, for various
    lattice sizes $L=2M+1$.  The data appear to converge toward a
    scaling curve in the large-$L$ limit, as demonstrated by the data
    reported in the inset for a particular value of the ratio
    $t/L^2$.}
  \label{rntn010ir}
\end{figure}

We now look for intermediate regimes of the time evolution, somehow
associated with the time scales of the Hamiltonian (\ref{Hfree})
driving the unitary dynamics, and therefore to its gap $\Delta_H$,
i.e. the energy difference between the first excited state and the
ground state. When keeping the particle number $N_0$ fixed, $\Delta_H$
behaves asymptotically as $\Delta_H \sim L^{-2}$, corresponding to the
dynamic exponent $z=2$ of the vacuum-superfluid
transition~\cite{Sachdev-book}.

We want to check whether the out-of-equilibrium dynamics develops
  an intermediate regimes somehow controlled by the Hamiltonian
  driving of the Lindblad equation, whose intrinsic time scale is
  related to its gap, i.e. $t \sim \Delta_H^{-1}\sim L^2$, which is
  much smaller than the time scale $t\sim L^3$ characterizing the
  approach to the stationary large-time limit (of course, for
  sufficiently large $L$, and in particular in the large-$L$
  limit). As we shall see, a closer look at the time evolution
  provides a clear evidence of such an intermediate regime, which also
  requires a rescaling of the dissipation strength, thus it emerges
  only at small values of $w$.

In Fig.~\ref{rntn010ir} we show some results for the time dependence
of the particle number in the case of systems with central-site
dissipation starting from a fixed number of particles. We observe that
the above-mentioned intermediate regime exists, and extends to any
$t\sim L^2$, if we perform an appropriate rescaling of the dissipation
parameter $w$, decreasing $w$ as $w\sim L^{-1}$. Indeed the numerical
results clearly support the large-$L$ scaling behavior
\begin{eqnarray}
 R_N(t,w,L) \approx U(t/L^2,wL)\,,   \label{scalbehwl}
%%  L^2 {d R_N(t,w,N_0,L)\over dt} &\approx& W(t/L^2,wL)\,,\nonumber
\end{eqnarray}
which corresponds to another FSS limit, obtained by increasing
  $L$ keeping the ratio $t/L^2$ and the product $wL$ fixed. Note that
this intermediate scaling behavior is expected to hold even for large
values of the ratio $t/L^2$, because it is also compatible with the
alternative scaling behavior (\ref{deltaLcenter}) of the Liouvillian
gap.

\subsection{Intermediate dynamic behavior keeping $N_0/M$ fixed}
\label{N0oLfixed}

\begin{figure}[!t]
  \includegraphics[width=0.95\columnwidth]{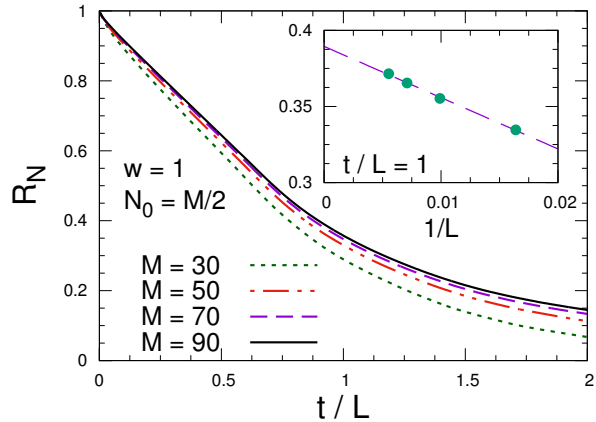}
  \caption{Behavior of the ratio $R_N$ versus $t/L$, for systems
    within hard walls with $N_0/M=1/2$, and central dissipation with
    $w=1$. The inset shows the $1/L$ approach to the large-$L$
    asymptotic behavior.}
  \label{rntn0lvstol}
\end{figure}

We now consider the large-$L$ behavior in the case we keep the ratio
$N_0/M$ fixed when increasing $L=2M+1$. This corresponds to the
superfluid phase, i.e. when the chemical potential $\mu$ is larger
than that at the vacuum-superfluid transition, $\mu>-2$. Within the
superfluid phase, the gap $\Delta_S$ of isolated free Fermi gases
behaves as~\cite{Sachdev-book} $\Delta_S \sim L^{-1}$.

Again we want to check whether the out-of-equilibrium dynamics in
  this condition develops an intermediate regime controlled by the
  part of the Lindblad equation driving the unitary dynamics
  introduces a time scale $t\sim \Delta_S^{-1}\sim L$ (again, much
  smaller than the time scale $t\sim L^3$ characterizing the approach
  to the stationary large-time limit). Like the case at fixed $N_0$,
  we show that such an intermediate regime exists, without tuning $w$
  toward zero.

The existence of a corresponding intermediate regime of the dynamics
is demonstrated by the results shown in Fig.~\ref{rntn0lvstol},
leading to the intermediate FSS ansatz
\begin{eqnarray}
 R_N(t,w,L) \approx W(t/L,w)\,,   \label{scalbehnol}
%%  L {d R_N(t,w,L)\over dt} &\approx& W(t/L,w)\,,\nonumber
\end{eqnarray}
obtained keeping $t/L$ fixed in the large-$L$ limit.  

\begin{figure}[!t]
\includegraphics[width=0.95\columnwidth]{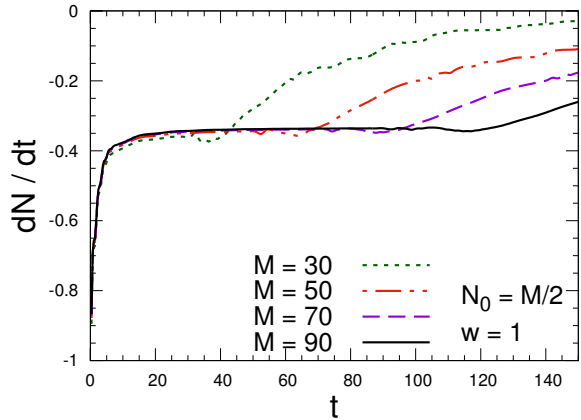}
\caption{The time dependence of the derivative of the particle number,
  for systems within hard walls with central-site dissipation with
  $w=1$, starting from ground states with $N_0/M=1/2$ fixed.  }
  \label{rntln0lvst}
\end{figure}

We finally report the existence of a further early-time regime when we
start from the ground state for $N_0\propto M$, as already noted in
Ref.~\cite{FMKCD-20}. Indeed, for sufficiently small time
\begin{equation}
  {d N(t,w,L)\over dt} \approx  f(t,w)\,,
  \label{vereg}
  \end{equation}
without showing any asymptotic size dependence.  This is shown by the
curves reported in Fig.~\ref{rntln0lvst}.  The behavior (\ref{vereg})
is observed in the large-size limit, and can be considered as the {\em
  thermodynamic} limit of the quantum evolution, when the time is
sufficiently small that the dynamics does not yet detect the effects
of the boundaries.  Indeed deviations are observed for $t\propto M$,
thus later and later for larger and larger systems. At the end of this
early-time regime, the intermediate regime (\ref{scalbehnol}) begins.

\section{Fermionic gases within harmonic traps}
\label{hartra}

We now present results for lattice fermionic systems within traps
arising from inhomogeneous external potentials, such as those in
Eq.~(\ref{potential}), in the presence of a particle-decay dissipation
at the center of the trap, as described by the Lindblad equations
(\ref{EQLindblad}) and (\ref{Lindop}). As already mentioned in
  the introduction, effective harmonic trapping mechanisms are quite
  common in cold-atom experiments~\cite{BDZ-08}. Therefore their
  analysis is also relevant from a phenomonological point of view.

We study the time evolution in the limit of large trap size $L_t$, in
the case we keep the initial particle number $N_0$ fixed, and when we
keep the ratio $N_0/L_t$ constant (equivalent to addiing a chemical
potential). We consider harmonic traps, thus $p=2$ in
Eq.~(\ref{potential}). The results are obtained for sufficiently large
systems $L$ at fixed $L_t$, so that a further increases of $L$ does
not change the results at fixed $L_t$, and therefore they can be
considered as results for infinite-size systems with a large accuracy,
within the accuracy of the numerical calculations, better than
$10^{-8}$.

\subsection{Large-time behavior}
\label{asytrap}

\begin{figure}[!t]
  \includegraphics[width=0.95\columnwidth]{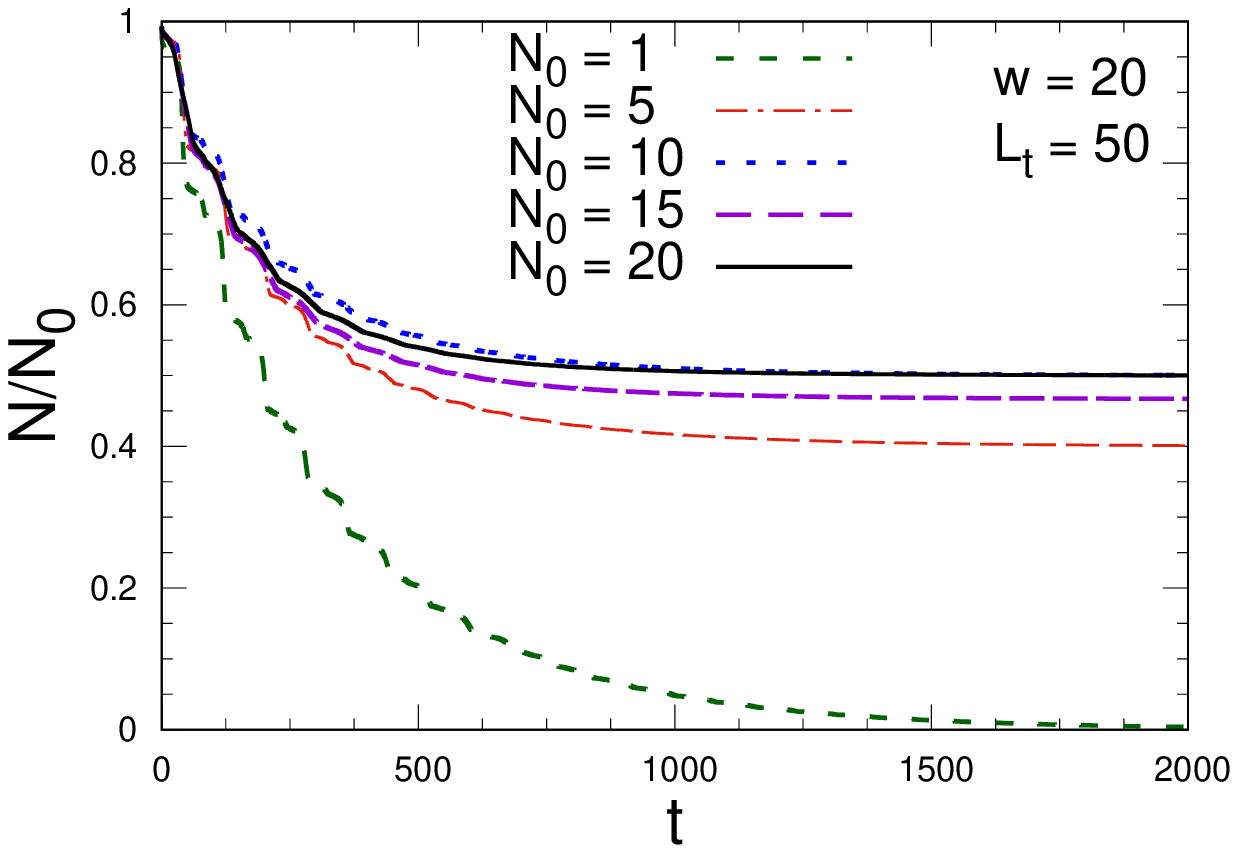}
\includegraphics[width=0.95\columnwidth]{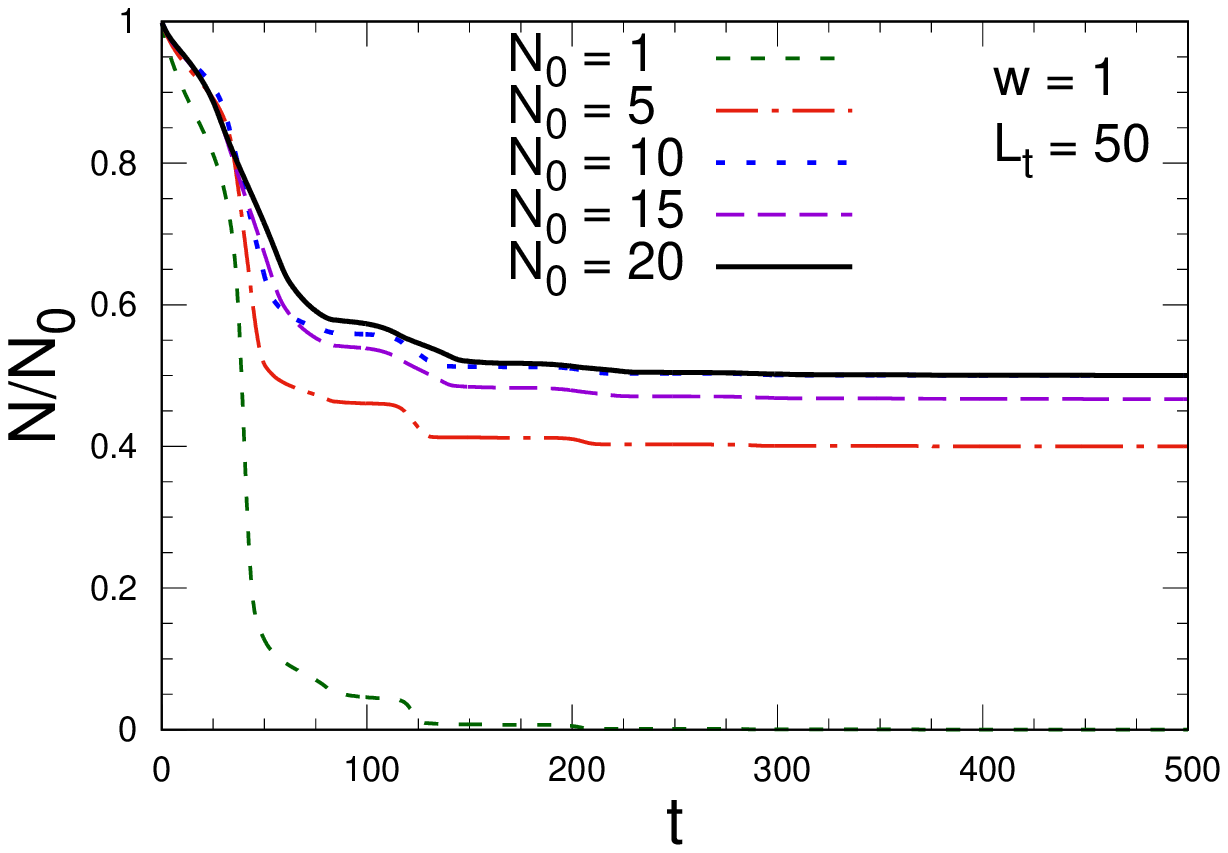}
\caption{The time dependence of the ratio $N/N_0$ for the central-site
  particle-loss dissipation $w=1$ (bottom) and $w=20$ (top) in systems
  within harmonic traps, various $N_0$, and $L_t=50$ (in the large-$L$
  limit to make the finite-size effects negligible).  The asymptotic
  stationary limit turns out to converge to $N/N_0=1/2$ for even
  $N_0$, and $N/N_0=(N_0-1)/(2N_0)$ for odd $N_0$. Similarly to the
  case of systems within hard walls, see Fig.~\ref{ndiffn0}, the
  approach to the asymptotic value turns out to be slower for $w=20$
  than $w=1$.}
\label{traptdep}
\end{figure}

To begin with, we discuss the asymptotic stationary states.  In
Fig.~\ref{traptdep} we show the time dependence, and asymptotic
behavior, of the ratio $N(t)/N_0$ for various values of the initial
number of particles. Again, similarly to the case of homogeneous
systems within hard walls and particle-decay dissipation at the center
of the chain, we find that the large-time states keep a half of the
initial particles.  Analogously to the hard-wall case, this can be
related to the fact that the one-particle Hamiltonian is invariant
under reflections with respect to the center, thus the one-particle
states must have definite parity. This can be easily seen in the
continuum limit, see e.g. Ref.~\cite{ACV-14}, where the one-particle
Hamiltonian eigenfunctions can be written in terms of Hermite
polynomials, and have definite parity.  Since the one-particle states
with negative parity vanishes at the center of the trap, the
corresponding modes in the ground state of the fermionic system are
not affected by the particle-decay dissipation at the center of the
trap. Then, recalling that the ground state is obtained by filling the
first $N_0$ one-particle levels, a selection mechanism analogous to
that identified in the case of homogeneous systems applies, see
Sec.~\ref{asysta}, therefore half of them are odd [more precisely
  $N_0/2$ for even $N_0$ and $(N_0-1)/2$ for odd $N_0$], we expect
again that half particles survive the central particle loss.

\begin{figure}[!t]
\includegraphics[width=0.95\columnwidth]{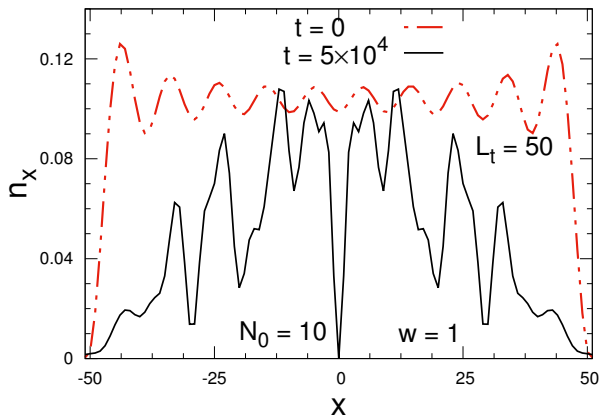}
\caption{The particle density $n_x$ for systems within a harmonic trap
  with initial particle number $N_0=10$, at $t=0$ and after some time
  $t$, for a central dissipation with $w=1$, fixed trap size $L_t=50$,
  and for a sufficiently large size of the system, $L=221$, to make
  finite-size effects negligible (numerically checked by verifying the
  dependence on $L$).}
\label{trapnxt}
\end{figure}

\begin{figure}[!t]
\includegraphics[width=0.95\columnwidth]{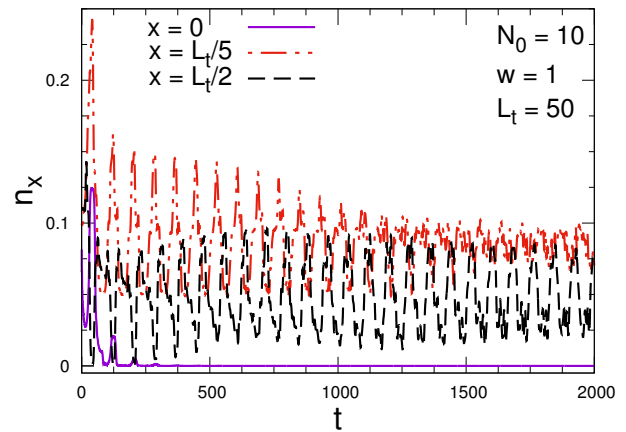}
\caption{ Time dependence of the particle density for systems within a
  trap of size $L_t=50$, at various spatial coordinates, $x=0$,
  $x=L_t/5=10$ and $x=L_t/2=25$, for central dissipation with $w=1$.
  The particle density $n_x$ for $x\neq 0$ are characterized by
  oscillations in the large-time regime, when $n_x$ at $x=0$ and the
  particle number have already approached their asymptotic behavior.
}
\label{trapnxto}
\end{figure}

In Fig.~\ref{trapnxt} we show the particle density at $t=0$ and at
large time when the central particle density has already stably
vanished. However, the spatial dependence of the particle density does
not appear static in the large-time regime where the particle number
stops decreasing. Indeed, as shown in Fig.~\ref{trapnxto}, the time
dependence of the particle density at sites $x\neq 0$ is characterized
by time oscillations that apparently continue indefinitely, persisting
even in the large-$L_t$ limit.  We also note that in this large-time
regime the quantum evolution of the particle density appears
effectively driven by the Hamiltonian term only. We have checked that
the dissipative contribution in Eq.~(\ref{eqscxytrap}), i.e. the one
proportional to $w$, gets suppressed asymptotically, thus only the
Hamiltonian determines the large-time dependence of the fixed-time
two-point function ${\mathscr C}_{xy}(t)$ and $n_x={\mathscr
  C}_{xx}(t)$. In this large-time regime also the energy of the system
defined in Eq.~(\ref{enedef}) remains constant, indeed its time
derivative vanishes when ${\rm Re}\,{\mathscr C}_{0,1}=0$,
cf. Eq.~(\ref{derene}).

\subsection{Large-$L_t$ scaling behavior of the time dependence}
\label{trapscaling}

\begin{figure}[!t]
\includegraphics[width=0.95\columnwidth]{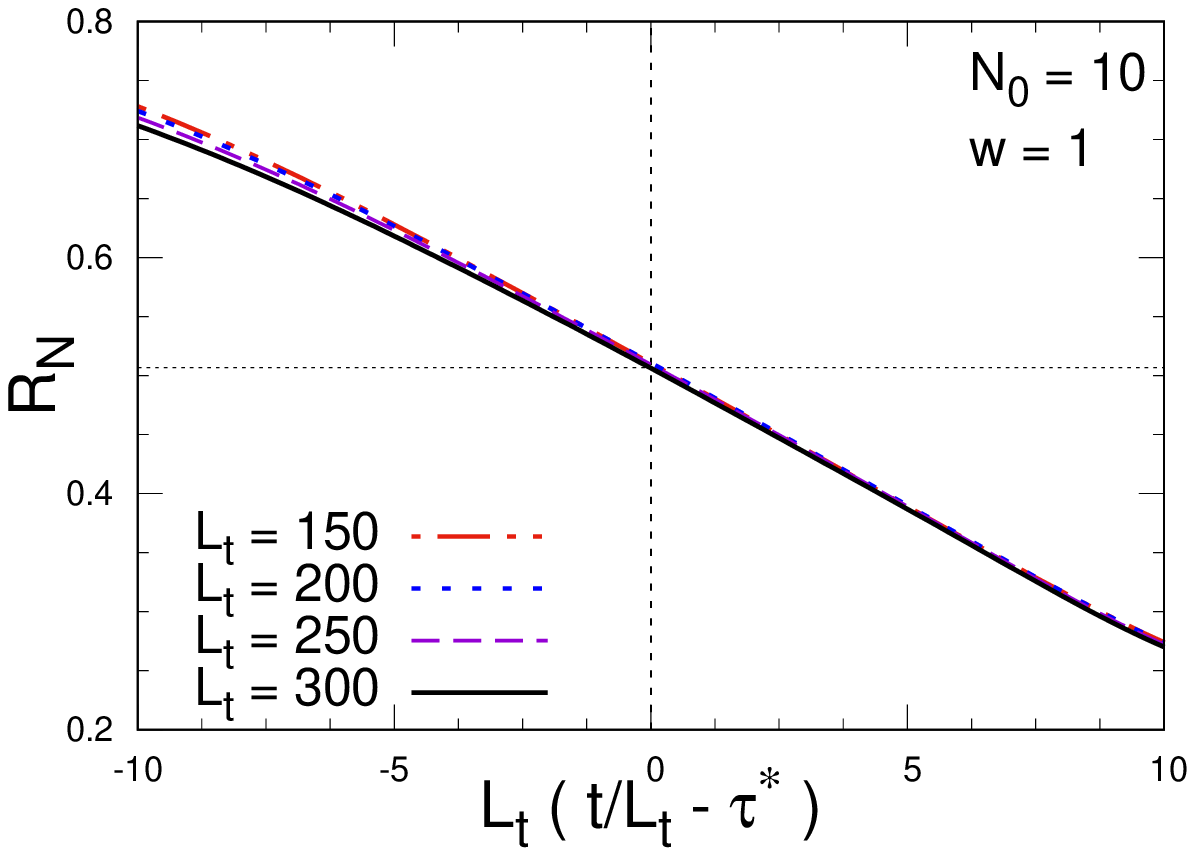}
\includegraphics[width=0.95\columnwidth]{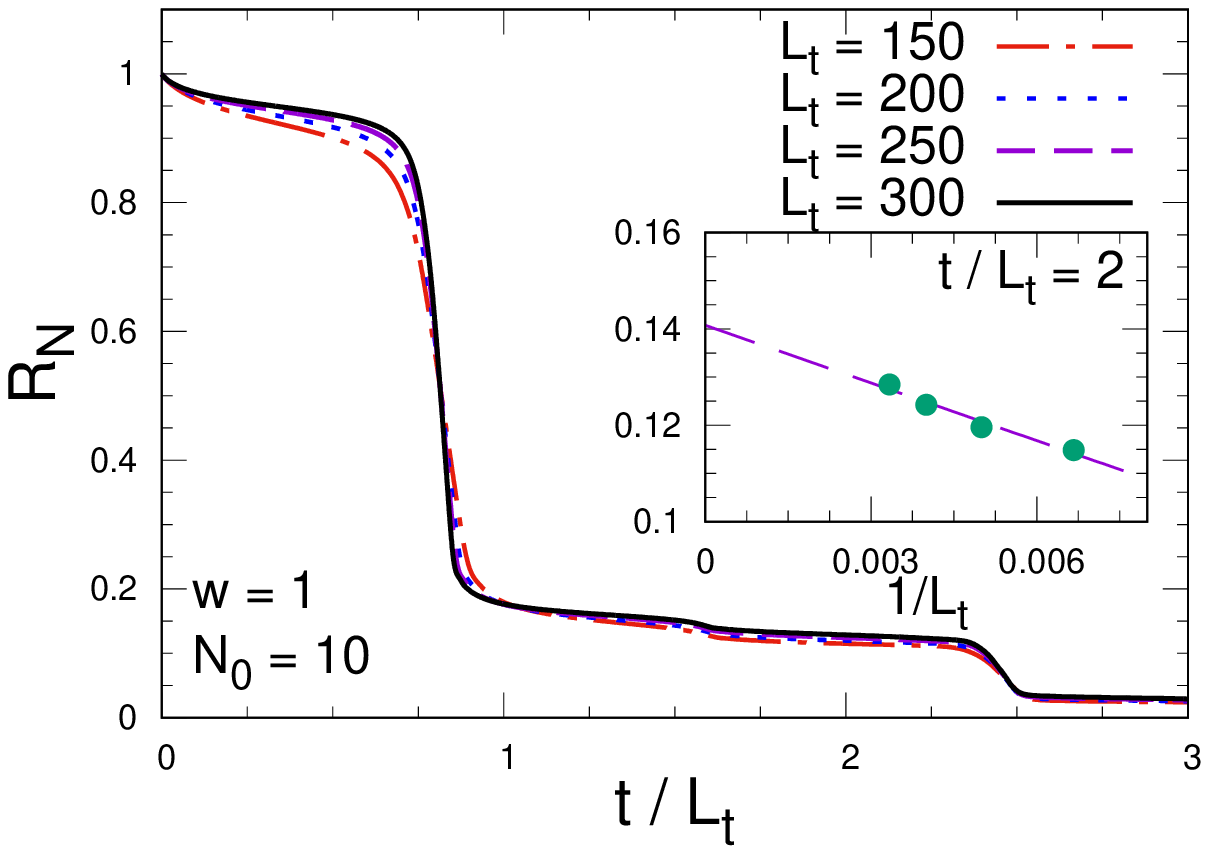}
\caption{ The bottom figure shows the time evolution of the ratio
  $R_N$ versus $t/L_t$ for $w=1$, various trap sizes, keeping the
  initial number of particles $N_0=10$ fixed.  The inset shows an
  example of convergence at $t/L_t=2$.  The top figure shows the
  scaling behavior at the fast drop of the particle number, around
  $t/L_t\approx 0.8147$, described by Eq.~(\ref{atcrit}).  }
\label{trapNo}
\end{figure}

\begin{figure}[!t]
\includegraphics[width=0.95\columnwidth]{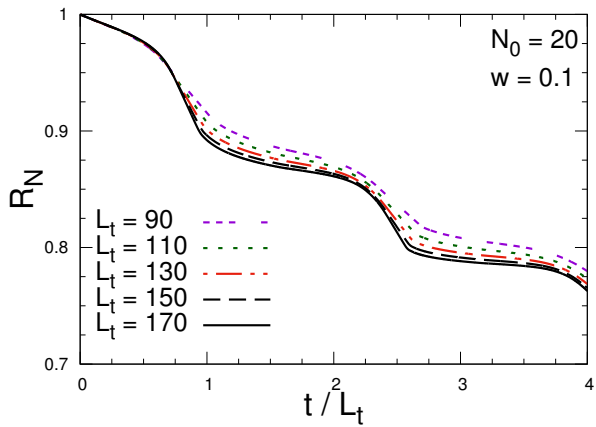}
\caption{The ratio $R_N$ for $w=0.1$, various trap sizes, keeping the
  initial number of particles $N_0=20$ fixed.}
\label{trapNow01}
\end{figure}

The time dependence starting from a fixed number of particles shows
the peculiar scaling behavior
\begin{equation}
  R_N(t,w,L_t) \equiv {N(t)-N_{\rm asy}\over N_0-N_{\rm asy}} \approx
  A_t(t/L_t,w)\,,
  \label{rntrapn0}
\end{equation}
which apparently describes the whole time evolution.  This is clearly
supported by the data reported in Figs.~\ref{trapNo} and
\ref{trapNow01} for central dissipation with $w=1$ and initial
particle number $N_0=10$.

Note that the time scale $t\sim L_t$ may be also related to the gap of
the fermionic Hamiltonian, for which $\Delta_{L_t} \sim
L_t^{-z\theta}$ where $z=2$ is the dynamic exponent associated with
the vacuum-to-superfluid transition, and $\theta=1/2$ is the universal
trap exponent characterizing critical behaviors in the presence of
trapping potentials~\cite{CV-09,CV-10,ACV-14,RV-21-rev} [for generic
  power laws of the potential (\ref{potential}), $\theta=p/(p+2)$].

\begin{figure}[!t]
\includegraphics[width=0.95\columnwidth]{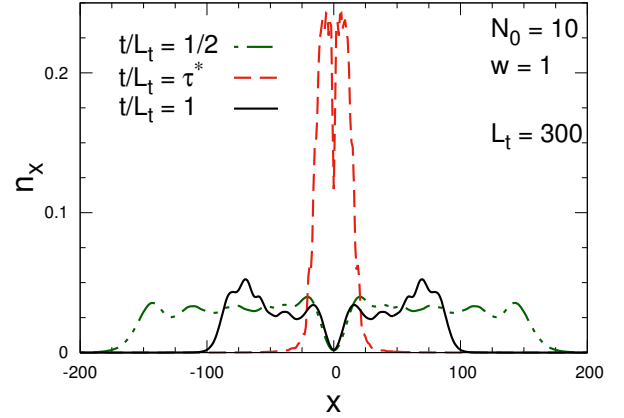}
\caption{ Behavior of the particle density around the singularity of
  the scaling function (\ref{rntrapn0}), for $L_t=300$, central
  particle loss with $w=1$, and some values of the ratio $t/L_t$
  around the singular point $t/L_t =\tau^*\approx 0.8147$. They show
  that large drop of the particle number at $\tau^*$ is connected with
  a simultaneous large increase of the particle density at the center
  of the trap.  }
\label{trapnx}
\end{figure}

The time scale $t\sim L_t$ characterizes the whole evolution of the
system, up to the large-time regime, except for some small
intermediate time intervals. Indeed, the curves shown in the bottom
Fig.~\ref{trapNo} presents some flat regions followed by rapid
changes.  Actually a more careful analysis, see, e.g., the top
Fig.~\ref{trapNo}, shows that the scaling function $A_t(t/L_t,w)$
entering Eq.~(\ref{rntrapn0}) appears to develop a singularity in the
large-time limit, at $t/L_t = \tau^*\approx 0.8147$, so that
\begin{equation}
  A_t(t/L_t,w) \approx f[L_t (t/L_t-\tau^*)]
  \label{atcrit}
\end{equation}
around $t/L_t=\tau^*$. Therefore this sharp drop of the particle
density occurs at a time $t^* \approx L_t \tau^*$ in the large-$L_t$
limit, and lasts for a finite time interval, i.e. it does not diverge
when increasing $L_t$.  Some data for the behavior of the particle
density and number around the time $\tau^*$ are shown in
Fig.~\ref{trapnx}, where we note the significant increase of the
particle density around $x=0$ when the singular behavior of the
scaling function (\ref{rntrapn0}) appears.  Analogous behaviors are
observed for generic values of $N_0$ and $w$.

\begin{figure}[!t]
\includegraphics[width=0.95\columnwidth]{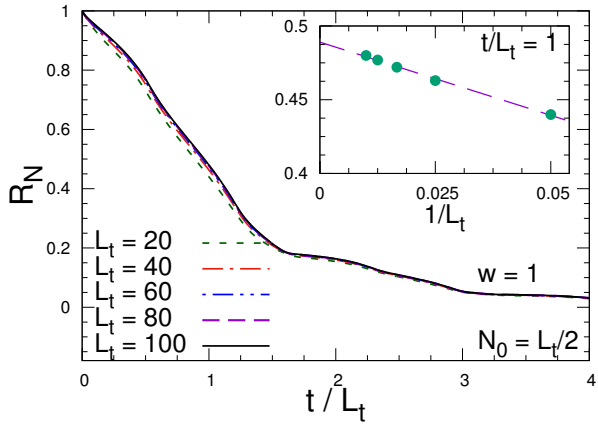}
\caption{ Time evolution of the ratio $R_N$ versus $t/L_t$ for
  fermionic gases within harmonic traps of various size, keeping the
  ratio $N_0/L_t=1/2$ fixed, for central particle-loss dissipation with
  $w=1$. The  large-$L_t$ convergence is evident, as also shown by
  the plot reported within the inset. }
\label{trapNoLt}
\end{figure}

In the case we keep $N_0/L_t$ fixed, the results show a generic
scaling behavior in terms of $t/L_t$, analogous to
Eq.~(\ref{rntrapn0}), see for example Fig.~\ref{trapNoLt}.

\section{Summary and conclusions}
\label{conclu}

We have addressed the out-of-equilibrium dynamics of fermionic gases
confined within finite spatial regions, subject to localized
dissipative interactions with the environment, entailing a localized
loss of particles.  We consider systems where particles are
constrained within a limited spatial region by hard walls, and systems
where they are trapped by a space-dependent potential, such as an
effective harmonic potential.  These issues are particularly
  relevant for cold-atom experiments, where atoms are confined within
  a limited spatial region by external potentials~\cite{BDZ-08}.
  Within this class of confined particle systems, we investigate the
  dynamic features arising from localized particle-loss dissipative
  mechanisms, which may be controllable, or inevitably present, in the
  experimental setup.  We discuss the quantum evolution at small and
  large times, focussing on the effects of finiteness of the confining
  region, which becomes eventually relevant for sufficiently large
  times.

As paradigmatic models we consider one-dimensional non-interacting
spinless fermionic lattice gases within hard-wall and harmonic traps,
with a particle loss localized at one of the sites of the chain, for
example at the center of the system, as sketched in
Fig.~\ref{fig:sketch}.  The dissipative particle-decay mechanism is
modeled within the framework of the Lindblad master equation governing
the time evolution of the density matrix.  Such a dissipative quantum
dynamics is analyzed within protocols starting from the equilibrium
ground state of the fermionic gas, then evolving under the effect of
the particle-loss dissipation, as outlined in Sec.~\ref{dynprot}.

We mostly exploit dynamic FSS frameworks, based on the definition
  of appropriate large-size limits.  This approach allows us to
  characterize the dynamics in the presence of localized particle-loss
  dissipation, identifying different dynamic regimes related to
  different features of the dissipative system, whose time scales are
  associated with powers of the size $\ell$.

The interplay between the time dependence and the size $\ell$ of the
system ($\ell$ corresponds to $M$ where $L=2M+1$ for hard-wall
confinement, and the trap size $L_t$, cf. Eq.~(\ref{potential}), for
inhomogeneous potential traps) is studied for two different initial
conditions: (i) fixed number $N_0$ of initial particles; (ii) keeping
the ratio $N_0/\ell$ constant, corresponding to the {\em
  thermodynamic} limit of the initial fermionic gases at equilibrium,
in both hard walls and harmonic traps.  These protocols provide us
with complementary information on the actual dynamics of finite-size
systems with relatively large size.  The time dependence of the
particle number and density show various dynamic scaling regimes in
the large-size limit, and nontrivial asymptotic large-time behaviors.
Notable differences emerge between homogeneous systems within hard
walls and systems within inhomogeneous harmonic traps.

The quantum evolution, arising from the dynamic protocol that we
consider, leads to asymptotic trivial empty states for generic
locations of the localized dissipative interaction. However there are
notable exceptions, in particular when the particle loss 
is localized at the center of the system, in both hard-wall and
harmonic traps. In this case only a fraction of the particles
eventually disappears, while half of them survives.  This is
essentially related to the non-interacting nature of the fermionic
gases, so that the localized dissipative mechanism preserves the
particle modes associated with the one-particle wave functions that
vanish at the site where the particle loss occurs, see
Sec.~\ref{asysta}. Therefore, due to the invariance under inversion
with respect to the center of the system, half one-particle modes
vanish at the center of the trap, thus they do not suffer from
particle loss.  The particle density at the site of the dissipation
vanishes for sufficiently large time. When this happens the particle
number $N(t)$ stops decreasing, cf. Eq.~(\ref{ntnz}), then the
residual particle number remains constant, because the localized
dissipative mechanism cannot reduce the particle number anymore. This
shows that there are two main different dynamic regimes in the
presence of one central particle-loss defect: the quantum dynamics is
initially driven by the particle-loss dissipation until the central
particle density gets asymptotically suppressed, then the subsequent
large-time evolution conserves the particle number and energy.

In the case of non-interacting fermionic gases confined within
  hard walls, of size $L=2M+1$, the out-of-equilibrium evolution
  arising from the protocol considered shows various dynamic regimes,
  which can be effectively distinguished by relating them to dynamic
  FSS limits corresponding to different time scales, which can be
  associated with the gap of the Liouvillian gap of the Lindblad
  equation, and the gap of its Hamiltonian driving. The main features
  of the out-of-equilibrium evolution can be summarized as follows.

(i) The asymptotic large-time states are characterized by a residual
number $N_{\rm asy}=N(t\to\infty)$ of particles: $N_{\rm asy} = N_0/2$
for even $N_0$, and $N_{\rm asy} = (N_0-1)/2$ for odd $N_0$.  Also the
particle density $n_x(t)$ shows a large-time stationary (time
independent) condition, with an almost constant $n_x$, except at $x=0$
where it vanishes.

(ii) The approach to the asymptotic stationary states is controlled by
the Liouvillian gap (\ref{deltadeb}), which decreases as $L^{-3}$
asymptotically. Therefore, for generic dissipative couplings $w$, the
times scale of the approach to the stationary state behaves as
$t_a\sim L^3$. Correspondingly, the particle-number ratio $R_N$
defined in Eq.~(\ref{defqu}) shows the asymptotic scaling behavior
$R_N(t,w,L)\approx A(t/L^3,w)$.

(iii) When we consider the large-size limit keeping the initial number
$N_0$ of particle fixed, the quantum evolution shows another
intermediate dynamic regime, i.e. $R_N\approx U(t/L^2,wL)$ which is
obtained in the large-$L$ limit by keeping $t/L^2$ and $wL$ fixed.
This regime may be associated with the gap of the Hamiltonian
  driving, which behaves as $\Delta_H\sim L^{-2}$.

(iv) Different intermediate regimes are observed when we keep $N_0/M$
fixed (equivalent to $N_0/L$ fixed) in the large-size limit.  We
observe an intermediate dynamic scaling behavior $R_N\approx
W(t/L,w)$, which can be related to the size-dependence $\Delta_s\sim
1/L$ of the Hamiltonian gap when the ratio $N_0/L$ is kept
fixed.

(v) When $N_0/M$ is kept fixed, there is also a further early time
regime, where the particle-number derivative depends only on time,
without showing any dependence on the size~\cite{FMKCD-20}. This may
be considered as the {\em thermodynamic} limit of the dynamics arising
from the protocol. Of course, this early-time regime stops when the
particle dynamics starts becoming sensitive to the boundaries,
i.e. for times $t\sim L$.

Substantially different behaviors emerge when we consider fermionic
gases trapped by an inhomogeneous harmonic potential within a region
of size $L_t$, cf. Eq.~(\ref{potential}).  We again consider the
dynamic in the case of one central particle-loss defect, whose main
features can be summarized as follows.

(i) The large-time behavior is again characterized by a residual
number of particles, analogously to the hard-wall case. The large-time
behavior of the particle density is again characterized by the
vanishing of $n_x$ at the dissipation site $x=0$, so that the particle
number stops decreasing. However the behavior of particle density
$n_x$ at $x\neq 0$ is not static in this large-time regime. Indeed it
shows sizable oscillations in time for $x\neq 0$, without apparently
reaching an asymptotic static condition, at least for finite trap
sizes.  We also note that, when the particle density at the central
site reaches its asymptotic zero value in the large-time regime, the
time evolution becomes effectively unitary for the particle density
$n_x$, i.e. it matches the behavior due to the only unitary driving in
the Lindblad equation, and conserves the energy.

(ii) The approach to the large-time behavior where $N(t)$ stops
decreasing is essentially characterized by time scales $t_a\sim L_t$.
This gives rise to the dynamic scaling behavior $R_N \approx
A_t(t/L_t,w)$ in the large-$L_t$ limits, in both cases when keeping
$N_0$ or $N_0/L_t$ fixed.  We also mention that, in the case of a
fixed initial particle number $N_0$, the scaling function
$A_t(t/L_t,w)$ shows intervals of $t/L_t$ with an almost flat (slowly
decreasing) dependence and apparent singularities between them, where
the particle number drops significantly in a small time scale
independent of $L_t$.

A natural extension of the study reported here may be that of
considering interacting fermionic systems, such as the Hubbard model.
Some results for interacting homogeneous fermionic systems have been
already reported in Refs.~\cite{FCKD-19,FMKCD-20,WSDK-20}, mostly
discussing infinite-size properties. Analogously to what done in this
paper for free fermionic gases, we believe that it is worth studying
the complex dynamic phenomena arising from the presence of the
boundaries, or inhomogeneous external trapping potential, even in the
case of interacting particle systems. Of course, the analysis of
interacting systems becomes significantly more challenge. Numerical
  methods to investigate interacting particle systems
  may be based on the so-called quantum trajectory algorithms,
which sample different quantum trajectories driven by an effective
Hamiltonian and stochastic quantum jumps, see e.g.
Refs.~\cite{KWBVBKW-15,FMKCD-20}.  Other possible extensions of our
work may consider different types of localized dissipative mechanisms,
and also higher dimensional systems.

We finally remark that the comprehension of the interplay between the
time dependence and the finite size of the confined system is
essential to interpret the behavior of physical systems when boundary
effects become important.  For example this issue is relevant for
small-size quantum simulators operating on a limited amount of quantum
objects, in the presence of controlled dissipation. We also mention
experiments with cold atoms within a trap of finite size, when the
many-body correlations become eventually sensitive to the
inhomogeneity arising from the trapping potential.

\appendix

\section{Liouvillian gap}
\label{appa}

To compute the Liouvillian gap for systems within hard walls, we
follow the method outlined in Ref.\cite{Prosen-08}. As explained
there, the results are obtained by diagonalization of the $4L\times
4L$ anti-symmetric complex matrix $A$ with nonzero elements
($\,j,\,k=-2M-1,\dots,2M\,$ and $L=2M+1\,$)
\begin{eqnarray}
      	&&A_{2j-1,2k-1} =  -2 i H_{jk} -  D_{jk}/2 + D_{kj}/2 \,, \nonumber \\
      	&&A_{2j-1,2k} =  i D_{kj} \,, \nonumber \\
      	&&A_{2j,2k-1} =  - i D_{jk}  \,, \nonumber \\
      	&&A_{2j,2k} =  -2 i H_{jk} + D_{jk}/2 - D_{kj}/2 \,, 
\end{eqnarray}
where the dissipation $2L\times 2L$ matrix $D_{j, k}$ describes the effects of
localized particle-loss dissipation at the central site,
\begin{eqnarray}
      	D_{j, k} = l_j\,l_k^* \,,\qquad
	l_j = \frac{w}{2}\left(\delta _{j,-2} - i\,\delta _{j,-1} \right) \,,
\end{eqnarray}
and the $(2L,2L)$ matrix $H_{j,k}$ is associated with the Hamiltonian 
(\ref{Hfree}) written in terms of the Majorana fermion operators $\{v_j\}$,
\begin{eqnarray}
	\hat H &= & \sum _{jk} \hat v_j H_{jk}\hat v_k = - \frac{1}{4}
        \sum _{j=-M}^{M} \Bigl( i\,\hat v_{2j-1}\hat v_{2j} -
        \notag\\ &-&i\,\hat v_{2j}\hat v_{2j+1} + i\,\hat v_{2j-1}\hat
        v_{2j+2} + {\rm h.c.}\Bigl) \,,
\end{eqnarray}
with ($\,x=-M,\dots,M\,$):
\begin{equation}
	\hat v_{2x} = \hat c_x + \hat c_x^\dagger\,,\quad
	\hat v_{2x-1} = i( \hat c_x^\dagger - \hat c_x ) \,.
\end{equation}
If $\beta _j$ are the eigenvalues of the matrix $A$, then the
Liouvillian gap is given by
\begin{equation}
     	\Delta _{\cal L} = 2\,{\rm Min}\,[{\rm Re}\,\beta _{j}]\,.
\end{equation}

\end{document}